\documentclass[twocolumn]{aastex63}
\usepackage{newtxtext,newtxmath}
\usepackage[T1]{fontenc}
\usepackage{ae,aecompl}
\usepackage{graphicx}	% Including figure files
\usepackage{amsmath}	% Advanced maths commands
\usepackage{amssymb}	% Extra maths symbols
\usepackage{booktabs}	% Pretty tables for all!
\usepackage{url}
\usepackage[pdf]{graphviz}

\newcommand{\asini}{$a \sin{i}$}

\newcommand{\maelstrom}{\textsc{Maelstrom}}

\received{}
\revised{}
\accepted{}
\submitjournal{AJ}

\shorttitle{Maelstrom}
\shortauthors{Hey et al.}
\graphicspath{{./}{figures/}}

\begin{document}

\title{Forward modeling the orbits of companions to pulsating stars from their light travel time variations}

\correspondingauthor{Daniel Hey}
\email{daniel.hey@sydney.edu.au}

\author[0000-0003-3244-5357]{Daniel R. Hey}
\affil{School of Physics, Sydney Institute for Astronomy (SIfA) \\
The University of Sydney, NSW 2006, Australia}
\affil{Stellar Astrophysics Centre, Department of Physics and Astronomy \\
Aarhus University, DK-8000 Aarhus C, Denmark}

\author[0000-0002-5648-3107]{Simon J. Murphy}
\affil{School of Physics, Sydney Institute for Astronomy (SIfA) \\
The University of Sydney, NSW 2006, Australia}
\affil{Stellar Astrophysics Centre, Department of Physics and Astronomy \\
Aarhus University, DK-8000 Aarhus C, Denmark}

\author[0000-0002-9328-5652]{Daniel Foreman-Mackey}
\affil{Center for Computational Astrophysics, Flatiron Institute, 162 5th Ave, New York, NY 10010, USA}

\author[0000-0001-5222-4661]{Timothy R. Bedding}
\affil{School of Physics, Sydney Institute for Astronomy (SIfA) \\
The University of Sydney, NSW 2006, Australia}
\affil{Stellar Astrophysics Centre, Department of Physics and Astronomy \\
Aarhus University, DK-8000 Aarhus C, Denmark}

\author[0000-0003-2595-9114]{Benjamin J.S. Pope}
\altaffiliation{NASA Sagan Fellow}
\affil{Center for Cosmology and Particle Physics, Department of Physics \\
New York University, 726 Broadway, New York, NY 10003, USA}
\affil{Center for Data Science, New York University, 60 Fifth Ave, New York, NY 10011, USA}

\author[0000-0003-2866-9403]{David W. Hogg}
\affil{Center for Cosmology and Particle Physics, Department of Physics \\
New York University, 726 Broadway, New York, NY 10003, USA}
\affil{Center for Computational Astrophysics, Flatiron Institute, 162 5th Ave, New York, NY 10010, USA}
\affil{Center for Data Science, New York University, 60 Fifth Ave, New York, NY 10011, USA}
\affil{Max-Planck-Institut f\"{u}r Astronomie, K\"{o}nigstuhl 17, D-69117 Heidelberg}

\begin{abstract}

Mutual gravitation between a pulsating star and an orbital companion leads to a time-dependent variation in path length for starlight traveling to Earth. These variations can be used for coherently pulsating stars, such as the $\delta$ Scuti variables, to constrain the masses and orbits of their companions. Observing these variations for $\delta$~Scuti stars has previously relied on subdividing the light curve and measuring the average pulsation phase in equally sized subdivisions, which leads to under-sampling near periapsis. We introduce a new approach that simultaneously forward-models each sample in the light curve and show that this method improves upon current sensitivity limits - especially in the case of highly eccentric and short-period binaries. We find that this approach is sensitive enough to observe Jupiter mass planets around $\delta$ Scuti stars under ideal conditions, and use gravity-mode pulsations in the subdwarf B star KIC~7668647 to detect its companion without radial velocity data. We further provide robust detection limits as a function of the SNR of the pulsation mode and determine that the minimum detectable light travel time amplitude for a typical \textit{Kepler} $\delta$ Scuti is around 2~s. This new method significantly enhances the application of light travel time variations to detecting short period binaries with pulsating components, and pulsating A-type exoplanet host stars, especially as a tool for eliminating false positives.

\end{abstract}


\section{Introduction}
\label{sec:introduction}

The detection and characterization of binary systems has traditionally relied upon three types of measurements: radial velocities (RVs) from spectroscopy, eclipses from photometry, and astrometry from imaging or interferometry. The availability of long duration space-based photometry from the \textit{Kepler} and Transiting Exoplanet Survey Satellite (\textit{TESS}) missions \citep{Borucki2010Kepler, Ricker2014Transiting} has facilitated a fourth method: measuring the influence of binary motion on stellar pulsations. 

For a star in a binary system, orbital motion leads to a periodic variation in the path length traveled by light emitted from each star and arriving at Earth. If at least one star is pulsating, the change in path length causes the observed phase of the pulsation to vary over the orbit \citep{Shibahashi2012FM}. This can be analyzed via two complementary techniques operating in the frequency and time domains: (i) frequency modulation (FM), where peaks in the amplitude spectrum are split by the orbital period and their amplitude ratios depend upon the orbital geometry of the system \citep{Shibahashi2012FM, Shibahashi2015FM}; and (ii) phase modulation (PM), where phases are directly extracted from the light curve and compared to a theoretical model \citep{Murphy2015Deriving, Murphy2016Finding,Murphy2018Finding}. PM has been used for several classes of stars, the most successful of which being pulsars, whose highly precise millisecond variations enable the detection of rocky planets. Indeed, the first discovered exoplanet was identified from light travel time variations \citep{Wolszczan1992Planetary}, with several more following \citep{Sigurdsson2003Young,Suleymanova2014Detection, Starovoit2017Existence}.

\begin{figure}
    \centering
    \includegraphics{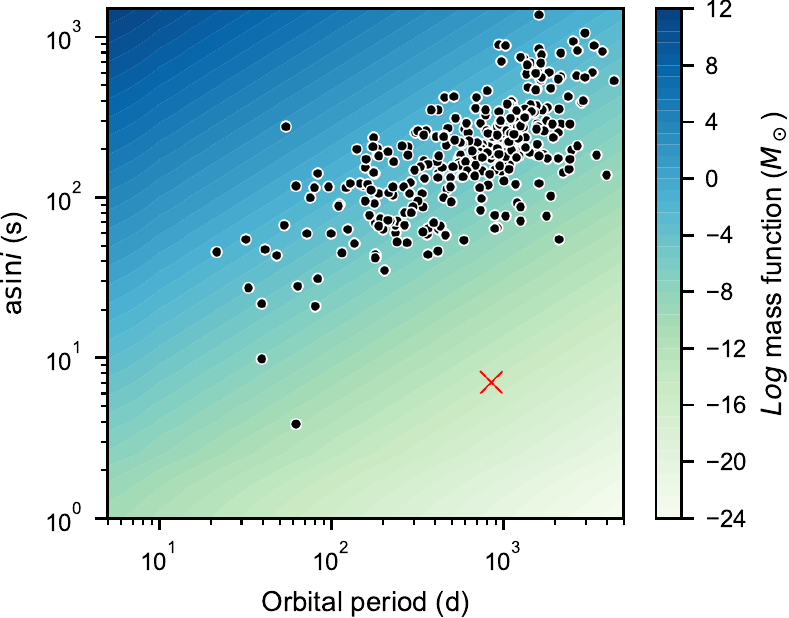}
    \caption{Distribution of binary systems discovered through PM via the subdividing method for the catalogue of \citet{Murphy2018Finding}. The mass functions are represented on a logarithmic scale. The red cross marks the only planet around a $\delta$~Sct star discovered through PM \citep{Murphy2016Planet}}
    \label{fig:known_binaries}
\end{figure}

For the intermediate-mass A-type $\delta$ Scuti stars, PM has been successful in identifying and characterizing more than 300 binary systems, most of which have intermediate orbital periods and mass ratios \citep{Murphy2018Finding}. By analogy with spectroscopic binaries, systems with one and two pulsating components are referred to as PB1 and PB2 respectively. Some of these systems have been found to have low mass ratios, consistent with planetary companions (Fig.~\ref{fig:known_binaries}). However, PM has been hampered in its efforts to uncover additional lower-mass companions by the signal-to-noise (SNR) of the pulsation frequencies \citep{Hermes2018Timing}.

In previous papers in this series \citep{Murphy2014Finding, Murphy2015Investigating, Compton2016Binary, Murphy2016Finding, Murphy2018Finding}, time-delay curves were extracted for pulsating binaries by subdividing the light curve (hereafter, the subdividing method). The measured time-delays were averaged over the corresponding subdivisions, which for highly eccentric binaries led to undersampling near periapsis. Although a correction for undersampling was formulated (\citealt{Murphy2016Finding}), orbits at high eccentricity are still significantly harder to detect. 

Here, we describe an approach that forward-models the PM effect directly on the light curve, mitigating the sampling problem. We provide worked examples of several well-studied systems and show that this method improves upon previous detection limits. We release a tested and open-source implementation of this method as a Python package, \maelstrom \footnote{\url{https://danhey.github.io/maelstrom/}}.

\section{New Approach}
\label{sec:new_approach}
We begin with an overview of phase modulation. In a binary system, the pulsating star's orbit around the barycentre leads to a change in path length for starlight traveling to Earth. The time-dependent flux variations due to $J$ pulsation modes therefore include a time-delay term, $\tau$:
\begin{multline}
  y(t) = \\ \sum_{j=1}^J \left[ A_j \cos (\omega_j[t-\tau(t)]) + B_j \sin(\omega_j[t-\tau(t)]) \right]  +  {\epsilon}
\label{eq:luminosity}
\end{multline}
where $A_j$, $B_j$ define the mode amplitudes, $\omega_j = 2\uppi\nu_j$ is the angular frequency of mode $j$, and $\epsilon$ describes variation unaccounted for by our pulsation model. We assume that $\epsilon$ is Gaussian distributed with variance $\sigma^2$ (encapsulating both measurement uncertainty and model mis-specification). Note that $\tau$ is defined to be positive when the star is on the far side of the barycentre, that is, farthest from the observer. 

Following \citet{Murphy2016Finding}, the time-delay can be expressed as a function of the true anomaly $f$:
\begin{eqnarray}
\tau = - \frac{a_1 \sin i}{c} \frac{1-e^2}{1+e \cos f} \sin(f+\varpi)
\label{eq:delay}
\end{eqnarray}
corresponding to a change in path length measured in light-seconds. We follow the previous convention that $a_1 \sin i$ denotes the projected semi-major axis of the pulsating primary, $e$ is the eccentricity, $\varpi$ is the angle between the ascending node and the periapsis, and $c$ is the speed of light.

In the subdividing method, the light curve is cut into equal subdivisions (typically 10~d in length), with the phase being measured in each division. A major strength of this approach is the independent calculation of the time delay for every oscillation mode in each division. This allows the response of each pulsation mode to the orbit to be visualized and checked for mutual agreement (e.g. figures~1--5 of \citealt{Murphy2014Finding}). A weighted-average time-delay is calculated across all modes in each division with the relative mode amplitudes $A_j$ and $B_j$ acting as weights, so that stronger modes are more heavily weighted. In our formulation here (Eq.\,\ref{eq:luminosity}), the mode amplitudes will also act as weights since the contribution of each mode to the flux variation is scaled by the amplitude of that mode, except that $\tau_j$ is now calculated at each individual observation $t$.

For the new formulation, we separate the time-delay equation (Eq.\,\ref{eq:delay}) into two components,
\begin{eqnarray}
\tau_{t, j} = \mathcal{A}_j \psi_t,
\label{eq:shape}
\end{eqnarray}
where
\begin{eqnarray}
\psi_t = -\frac{1-e^2}{1+e \cos f} \sin(f+\varpi),
\label{eq:psi}
\end{eqnarray}
and $\mathcal{A}_j$ is the projected semi-major axis evaluated for every mode, $j$,
\begin{eqnarray}
\mathcal{A}_j = a_j \sin{i} / c.
\label{eq:aj_comp}
\end{eqnarray}
Modes can be grouped together by their $\mathcal{A}_j$, which is particularly useful when there are two pulsators in the same binary system (PB2; e.g. figure~6 of \citealt{Murphy2014Finding}).

We construct the design matrix, $D_j$, for each $j$-th mode and each observation time $t_n$, and combine these into the master design matrix $D$, which has $N$ rows (for $N$ data-points) and $2J$ columns:

\begin{eqnarray}
D_j = \left( \begin{array}{cc}
	\cos (\omega_j[t_1 - \tau_{1,j}])& \sin (\omega_j[t_1 - \tau_{1,j}]) \\
	\cos (\omega_j[t_2 - \tau_{2,j}])& \sin (\omega_j[t_2 - \tau_{2,j}]) \\
	\vdots & \vdots\\
	\cos (\omega_j[t_n - \tau_{n,j}])& \sin (\omega_j[t_n - \tau_{n,j}]) \\
	\vdots & \vdots \\
	\cos (\omega_j[t_N - \tau_{N,j}])& \sin (\omega_j[t_N - \tau_{N,j}])
	\end{array}\right)
\label{eq:Dj}
\end{eqnarray}
and
\begin{eqnarray}
D = (D_1~D_2~\dots~D_j~\dots~D_J).
\label{eq:D}
\end{eqnarray}

The amplitude coefficients $A_j$ and $B_j$ are collected in the column matrices $w_j$
\begin{eqnarray}
w_j = \left( \begin{array}{c}
	A_j \\
	B_j
	\end{array} \right),
\label{eq:DW}
\end{eqnarray}
which are combined into the matrix of weights, $w$:
\begin{eqnarray}
w = \left( \begin{array}{c}
	w_1 \\
	w_2 \\
	\vdots \\
	w_j \\
	\vdots \\
	w_J \\
	\end{array} \right).
\label{eq:w}
\end{eqnarray}
The variance in the light curve, $y$ (Eq.\,\ref{eq:luminosity}), is then expressed as
\begin{eqnarray}
\label{eqn:lightcurve}
y = D \cdot w + \epsilon.
\label{eq:lightcurve}
\end{eqnarray}
This theoretical light curve can be compared to the actual light curve as posed through an optimization problem. Calculating this requires the likelihood,
\begin{eqnarray}
\ell = -\dfrac{1}{2} \sum_{n} \frac{\left( y_n - (D\cdot w)_n  \right)^2}{\sigma^2},
\label{eq:logl}
\end{eqnarray}
where $\sigma$ is the standard deviation of each measurement, $y_n$.

The maximum likelihood values for the weights can be found by ordinary least squares linear regression:
\begin{eqnarray}
\hat w = (D^T \cdot D)^{-1}(D^T \cdot y_n),
\label{eq:What}
\end{eqnarray}
where $D^T$ is the transpose of $D$ and $y_n$ is the flux at time $t_n$. The value of the log-likelihood ($\ell$) at this maximum is thus Eq.~\ref{eq:logl} with $\hat w$ substituted for $w$. Note that this is an approximation to the log-likelihood in which we presume that $\hat w$ is so precisely determined that its uncertainty can be ignored. Despite that, it is worth pointing out that it is not necessary to marginalize over $\hat w$, and that it can be accounted for following \cite{Luger2017Linear}.

Our approach thus changes the problem from fitting extracted time-delays to instead comparing the entire light curve against a theoretical model. This is a similar approach to that implemented by \cite{Silvotti2007Giant} and \cite{Telting2012Three, Telting2014KIC}.

\section{Implementation}
\begin{figure}
    \centering
    \includegraphics[width=\linewidth]{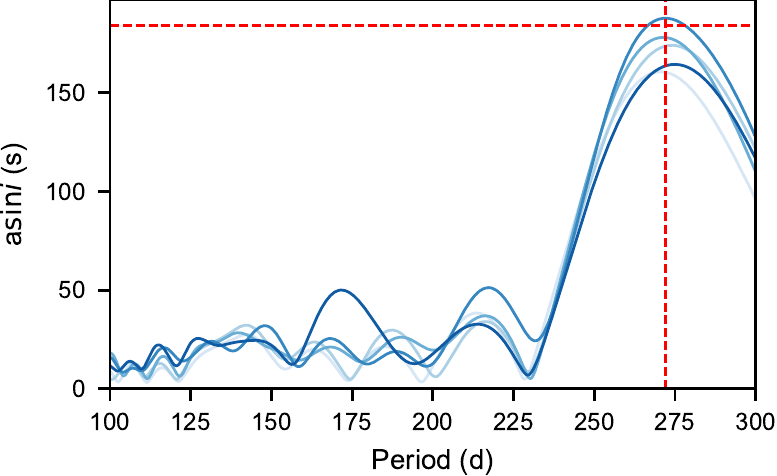}
    \caption{Result of optimising the model over a grid of orbital periods (between 100 and 300~d) for KIC~9651065. The model is fixed to each orbital period, and optimised to find the projected semi-major axis (\asini) for each mode J. We use the 5 strongest modes, with opacity proportional to their amplitude. The dashed red lines indicate the values of the system obtained in Sec~\ref{sec:9651065}.}
    \label{fig:9651065_period_search}
\end{figure}

We implement a Python package, \maelstrom\ \footnote{A static version of the code used throughout this paper is available through Zenodo: \cite{daniel_2020_3669395}.} to perform inference using the light curve model described above. \maelstrom\ is written with \textsc{PyMC3}, a probabilistic programming module for Bayesian inference \citep{Salvatier2016Probabilistic} designed with the \textsc{Theano} model building framework \citep{TheTheanoDevelopmentTeam2016Theano} and \textsc{exoplanet} \citep{DanForeman-Mackey2019Dfm}. \textsc{PyMC3} implements Markov Chain Monte Carlo (MCMC) methods through the No-U-Turn-Sampler \citep[NUTS;][]{Hoffman2011NoUTurn}, which scales well to high-dimensional and complex posterior distributions. \maelstrom\ provides functionality for fitting time-delays using both the subdividing and forward-model approaches. We provide a `general' model, which has pre-defined priors on the orbital parameters, as well as utilities for calculating Eq.~\ref{eqn:lightcurve} given arbitrary inputs, allowing for more tailored models and custom priors to be defined.

%CRAP: \textsc{PyMC3} implements a Hamiltonian Markov-Chain Monte-Carlo sampler (HMC), a form of gradient-based inference \citep{Neal2012MCMC}. The main limitation of gradient-based inference methods is that to use them, the gradients must be known and solved. The fundamental quantity of interest is the first derivative (gradient) of the log-likelihood with respect to the parameters of the model, which is not always possible to solve analytically. Instead, it is generally preferable to use a method known as automatic differentiation that can compute the exact gradients of the model to machine precision at compile time \citep{Baydin2015Automatic}. We use \textsc{Theano} for its support for auto differentiation.

%Solving the time-delay equation (Eq.~\ref{eq:delay}) further requires computing the transcendental Kepler equation and its derivatives which can not be obtained analytically. For this reason, we use the Kepler equation solver implemented in \texttt{Exoplanet} \citep{DanForeman-Mackey2019Dfm}.

Below we discuss some of the more subtle aspects of modeling time-delays, such as choosing pulsation frequencies, calculating the uncertainties, and including additional data constraints. While we strive to cover several varied examples, it is important to note that these should serve as a guideline only. Accompanying the \maelstrom\ code are several tutorials which discuss common pitfalls in modeling these systems and the considerations when choosing priors.

\subsection{Choice of pulsation frequencies}
\label{sec:implementation}
\begin{figure}
    \centering
    \includegraphics[width=\linewidth]{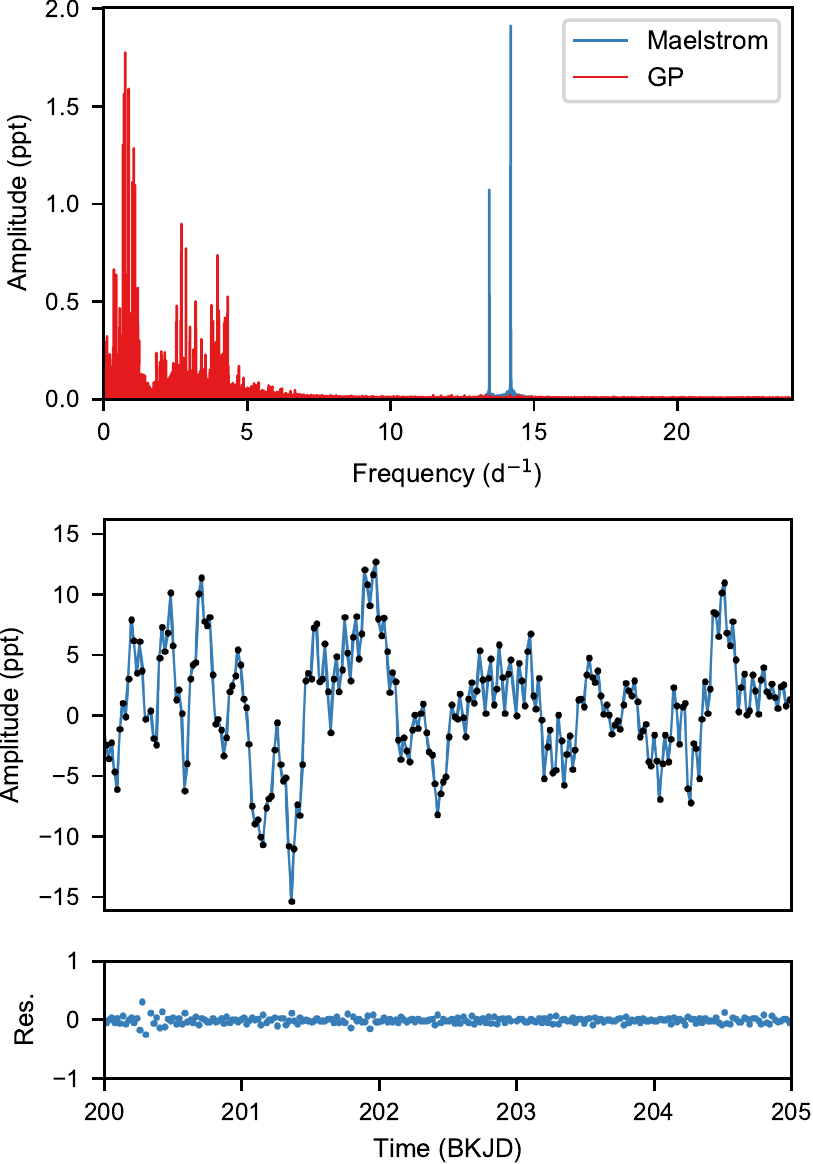}
    \caption{Simultaneous GP and time-delay fit to KIC~6780873, a PB1 system with significant low frequency variation. The top panel shows the amplitude spectrum of the light curve coloured by the contributions from the GP posterior prediction (red) and the time-delay model (blue). The middle panel shows the original light curve data (black points) against the model fit obtained after optimization (blue curve). There is no periodic variation in the residuals (bottom panel), indicating a good fit.}
    \label{fig:6780873_lc_model}
\end{figure}

Light arrival time variations must, by definition, affect all pulsation modes equally. However, not all pulsations in a star have usable time-delay information. Both mode crowding and the signal-to-noise ratio (SNR) influence whether an orbital signal can be observed. The stability of the pulsation mode itself also determines its usefulness. Stochastic oscillations as seen in solar-like stars and red giants are not suitable for PM. PM depends upon coherent pulsations, and in general, has only been applied to the pressure mode pulsations in $\delta$ Scuti stars \citep{Compton2016Binary}. Whilst white dwarfs and subdwarf B stars are promising `clocks' for this method, recent work has indicated that their pulsations are not as stable as $\delta$~Scutis and thus should be treated with caution when being used for PM, especially for longer period binaries \citep{Zong2018Oscillation}.

Several approaches can be used to determine which pulsation modes are useful for phase modulation. For longer period binaries, one can simply subdivide the light curve and inspect the phases of each frequency for orbital motion, as has been done in previous papers in this series. At shorter orbital periods this becomes more difficult, since the time-delay signal is smeared by the chosen segment size of the subdivision, and eventually, the period becomes shorter than the Nyquist limit imposed by the subdivisions. Instead, the model (Eq.~\ref{eq:luminosity}) is fitted to the light curve, with each frequency having an independent \asini\ (the $\mathcal{A}_j$ component of our formulation). The model can then be optimized and each \asini\ checked for mutual agreement. As an alternative, one can iterate over a grid of orbital periods while recording the log-likelihood and optimized \asini\ at each period for every mode. Modes with usable time-delay information will also demonstrate a mutual agreement in their optimized values, as seen for the binary system KIC~9651065 in Fig.~\ref{fig:9651065_period_search}.

There is an additional correction to the pulsation frequencies required due to the time-averaged Doppler shift of those frequencies over the orbit, which causes the measured pulsation frequencies to be slightly offset with respect to their true values (see, for example, figure~10 of \citet{Murphy2016Finding}). For an integer number of observed orbits, the frequency of pulsation calculated from the amplitude spectrum is the true one. However, if the observations instead span a non-integer number of orbits the pulsation frequency is shifted by an amount
\begin{eqnarray}
    \Delta \nu_{\rm osc}  = \dfrac{T\;\mathrm{mod}\;P_{\rm orb}}{T} \dfrac{\nu_{\rm osc}}{c} \int_{0}^{\rm T} \nu_{\rm rad}(t) dt,
    % I am only using <> for the average because i like the sound of langle and rangle
    %\Delta \nu_{\rm osc}  = -\dfrac{T\,\mathrm{mod}\, P_{\rm orb}}{T} \nu_{\rm osc} \langle\tau(t)\rangle
\end{eqnarray}
where $T$ is the time-span of the data. Previously, a single drift term $m$ was introduced to account for variation of the time-delays across the light curve. That is, the time-delays were scaled by a gradient $\tau(t)' = \tau(t) + mt$. In our formulation, we instead let each frequency $\omega_j$ in Eq.~\ref{eq:delay} be a free parameter in the model that does not vary with time. This removes the need to account for time-delay drift while simultaneously fitting the pulsation frequencies.

In addition to the pulsation frequencies used for the time-delay analysis, it is common for hot stars to exhibit low-frequency variations due to rotational modulation, subsurface convection, or low-frequency gravity mode (g-mode) pulsations \citep{Li2019Period,Bowman2019Lowfrequencya,Bowman2019Photometric,Cantiello2019Envelope}. When this occurs, there is a large discrepancy between the observed and calculated light curves. We model these signals with a Gaussian Process (GP), as implemented in \textsc{celerite} \citep{Foreman-Mackey2017Fast, Foreman-Mackey2018Scalable}, assuming that the low-frequency signals are described by a simple harmonic oscillator kernel. We show such a model in Fig.~\ref{fig:6780873_lc_model} fitting both the phase modulations in the light curve and a GP model for the low-frequency g-modes in a $\delta$~Sct binary, KIC~6780873. The orbital properties of this star are derived completely in Sec.~\ref{sec:6780873}.

\label{sec:choice}

\subsection{Radial velocities}

Radial velocities depend on the same orbital elements as time-delays. Indeed, the radial velocity $\nu_{\rm rad}$ is related to the time delay by,

\begin{eqnarray}
\nu_{\rm rad} = c \dfrac{d\tau}{dt},
\end{eqnarray}
where $c$ is again the speed of light. Differentiating Eq.~2 with respect to time, it is straightforward to obtain the radial velocity:
\begin{eqnarray}
\nu_{\rm rad} = - \dfrac{a_1 \sin{i}}{P_{\rm orb}}\dfrac{2\pi}{\sqrt{1 - e^2}}\big[\cos{(f + \varpi) + e\cos{\varpi}}\big].
\label{eqn:radvel}
\end{eqnarray}

There has been ambiguity in previous papers regarding the sign convention on the radial velocities and time-delays. The sign depends on the chosen orbital convention. Here, we assume $\varpi$ is measured from the ascending node, such that both the time-delay (Eq.~\ref{eq:delay}) and radial velocity (Eq.~\ref{eqn:radvel}) are preceded with minus signs. This ensures that objects have a positive radial velocity when moving away from us, and that a positive delay is experienced when light has a longer path length to traverse and thus arrives "late", as expected.

Time-delays can be directly converted into RVs, either to extend the coverage of separate RV measurements or predict them where none are available. On the other hand, orbits can be solved simultaneously with time-delays and RVs, adding an extra constraint for the model. We give examples of this in Sec.~\ref{sec:solution}.

\subsection{Calculating the uncertainties}

The forward-model is formulated under the assumption that uncertainties on the parameters will be obtained through MCMC. It is also possible to obtain lower bounds on the uncertainties using the Fisher information \citep{Ly2017Tutorial}. For a given parameter $\theta$ in the model whose maximum likelihood estimate is $\hat \theta$, the log-likelihood is expressed as

\begin{eqnarray}
\hat{\ell} = -\frac{1}{2}\frac{(\theta - \hat \theta)^2}{\sigma_{\theta}^2},
\end{eqnarray}

\noindent The variance $\sigma_{\theta}^2$ can be found by computing the partial derivatives of the likelihood at this estimate, in the Hessian matrix:
\begin{eqnarray}
H = -
\left( \begin{array}{cccc}
  \tfrac{\partial^2}{\partial \theta_1^2}
  &  \tfrac{\partial^2}{\partial \theta_1 \partial \theta_2}
  &  \cdots
  &  \tfrac{\partial^2}{\partial \theta_1 \partial \theta_p} \\
  \tfrac{\partial^2}{\partial \theta_2 \partial \theta_1}
  &  \tfrac{\partial^2}{\partial \theta_2^2}
  &  \cdots
  &  \tfrac{\partial^2}{\partial \theta_2 \partial \theta_p} \\
  \vdots &
  \vdots &
  \ddots &
  \vdots \\
  \tfrac{\partial^2}{\partial \theta_p \partial \theta_1}
  &  \tfrac{\partial^2}{\partial \theta_p \partial \theta_2}
  &  \cdots
  &  \tfrac{\partial^2}{\partial \theta_p^2} \\
\end{array} \right) 
{\hat \ell}
\label{eq:sigma}
\end{eqnarray}

\noindent which is a square matrix with $M$ rows and columns. 

For a given set of optimal parameters ($\hat \theta$), the standard errors are the square roots of the diagonal elements of the covariance matrix, defined as the inverse of the Hessian,

\begin{eqnarray}
\sigma_{\theta} = \sqrt{{\rm diag}(H^{-1})}.
\end{eqnarray}

This approach to calculating uncertainty is useful for determining the lower bound on error in the vicinity of known model parameters, and for uncertainty estimation in situations where the maximum-likelihood estimate is close to the true value. In general, proper sampling and exploration of the posterior distribution of the model yields the most accurate uncertainties.

\section{Examples}
\label{sec:examples}
\begin{table}
    \centering
    \caption{Priors chosen for both the subdividing and forward-models, where `estimated' refers to a value obtained from initial inspection of the data. Priors on the orbital period and semi-major axis have been forced positive.}
    \label{tab:priors}
    \begin{tabular}{cc}
        \hline
        {Parameter}    & {Prior}\\
        \hline
        $P_{\rm orb}$           & $\mathcal{N} \sim (P_{\rm estimated}, 5)$\\
        $\rm{a}\sin{i}/c$       & $\mathcal{N}\sim (\rm{a}\sin{i}/c_{\rm estimated}, 10)$\\
        $e$                     & $\mathcal{U}\sim (0,1)$  \\
        $\varpi$                & $\mathcal{U} \sim (0, 2\pi)$\\
        $\phi_p$                & $\mathcal{U} \sim (0, 2\pi)$      \\
        \hline
    \end{tabular}
\end{table}

To verify the new approach we applied it to several well-characterized systems that have previously been studied with PM. We compared the subdividing and forward-model approaches, and when available, radial velocities were used to further constrain the orbit. We used the Pre-search Data Conditioning Simple Aperture Photometry (msMAP; \citealt{Twicken2010Presearch, Smith2012KeplerPresearch, Stumpe2014Multiscalea}) \textit{Kepler} light curves downloaded from the Barbara A. Mikulski Archive for Space Telescopes (MAST). Times and orbital parameters are reported in the Barycentric Kepler Julian Date (BKJD) corresponding to BJD-2454833.

\label{sec:solution}
\begin{figure}
    \centering
    \includegraphics{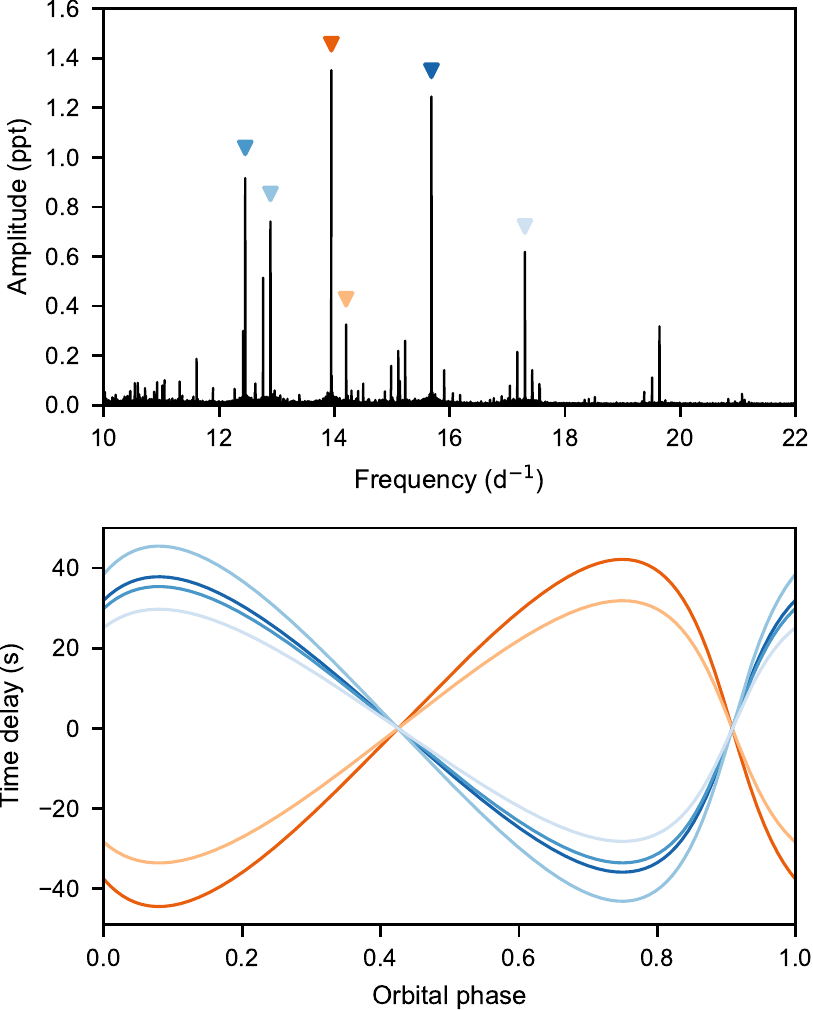}
    \caption{A binary system in which both stars are $\delta$~Sct pulsators, KIC~10080943. The top panel shows the amplitude spectrum with the seven highest amplitude pulsations chosen for time-delays. The bottom panel shows the result of optimising the forward-model for each of these frequencies simultaneously, where each frequency has an independent \asini\, folded on the orbital period of 15.3 d. Doing so allows for the pulsation modes to be assigned to each star, with blue and orange corresponding to stars A and B respectively. This system is modeled completely in Sec.~\ref{sec:10080943}.}
    \label{fig:10080943_init_opt}
\end{figure}

To obtain an orbital solution, we first identified pulsation frequencies that agree in orbital period and \asini\ for their time-delays. In the forward-model, each chosen frequency has an independent \asini\ that is allowed to be either negative or positive, corresponding to the primary and secondary stars, respectively. This is the amplitude component of our formulation, $\mathcal{A}_j$ (Eq.~\ref{eq:psi}). 

The model was then optimized to find the maximum a posteriori (MAP), using the L-BFGS-S routine \citep{Zhu1994LBFGSB} as implemented in SciPy \citep{Jones2001SciPy}. In the case where only one star in the system is pulsating (PB1), each \asini\ is positive, and the initial estimate was taken as the median of these values. When both stars are pulsating, individual pulsation frequencies can be assigned to the stars based on the sign of \asini. Pulsations belonging to different stars will have \asini\ of flipped sign. We show this for the case of the $\delta$~Scuti binary system KIC~10080943 in Fig.~\ref{fig:10080943_init_opt}.

The model was then re-created with an \asini\ of singular shape. That is, $\mathcal{A}_j$ was instead fixed to a single value per pulsating star instead of one independent \asini\ per pulsation frequency. This was done to account for the fact that each star in a binary only has a single measure of the projected semi-major axis, where each individual mode in $\mathcal{A}_j$ is simply an independent measure of it. The model was again optimized and the NUTS sampler was initialized on these optimized values, with priors as defined in Table~\ref{tab:priors}. We ran 1000 steps of burn-in and 2000 steps of MCMC over 2 chains simultaneously. To assess convergence, we inspected the integrated auto-correlation time of the chain, averaged across each parameter as an estimate of the number of effective samples. 

We report values of the orbital parameters at the median, with uncertainties taken from the $16$th and $84$th percentiles of their posterior distribution. We also include the code used to generate these models in the \maelstrom\ repository, which includes the final posterior distribution of each sampled parameter. For angular orbital parameters, such as the phase and angle of periapsis ($\phi_p$, $\varpi$), we used an angle prior that samples a unit vector from sine and cosine components simultaneously, ensuring that the sampler does not see a discontinuity at $-\pi$ or $\pi$.

Although we are modeling the actual light curve, we show derived time-delay curves as a convenient proxy to visualize the orbit. We find that in general, previously reported uncertainties using the subdividing method were underestimated, which we attribute to improper exploration of the parameter space. %When modeling the time-delays through the subdivided method, we used identical data from previous papers for consistency.

\subsection{KIC 9651065: PB1, PM}
\label{sec:9651065}

\begin{figure}
    \centering
    \includegraphics[width=\linewidth]{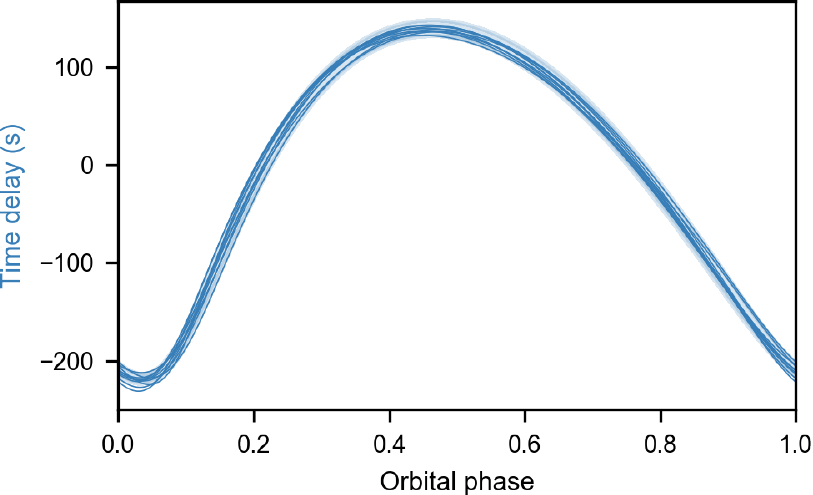}
    \caption{Derived time-delay (blue) model fit from the forward-model for KIC~9651065. Shown are 10 randomly selected models from the trace after burn-in. The shaded blue region shows the standard deviation of the posterior distribution.}
    \label{fig:9651065_td}
\end{figure}

{\renewcommand{\arraystretch}{1.3}%
\begin{table}
    \centering
    \caption{Fitted orbital parameters for KIC~9651065 through both the old and new methods. $f(M)$ is the derived mass function of the system, and is convolved with the (unknown) angle of inclination.}
    \label{tab:9651065}
    \begin{tabular}{crrc}
        \hline
            & {Subdividing} & {Forward} & {Unit}  \\
        \hline
        $P_{\rm orb}$           & $272.35^{+0.47}_{-0.49}$          & $272.27^{+0.42}_{-0.42}$      &   d\\
        $\rm{a}_1\sin{i}/c$       & $184.4^{+3.9}_{-3.7}$          & $185.2^{+3.5}_{-3.7}$      &   s\\
        $e$                     & $0.45^{+0.04}_{-0.03}$              & $0.45^{+0.03}_{-0.03}$        &   \\
        $\varpi$                & $-1.02^{+0.08}_{-0.08}$              & $-0.99^{+0.07}_{-0.07}$       &   rad    \\
        $\phi_p$                & $0.49^{+0.08}_{-0.08}$               & $0.52^{+0.07}_{-0.07}$       &   \\
%         $f(M)$                  & $0.091^{+0.007}_{-0.006}$       & $0.092^{+0.006}_{-0.005}$  &   $\mathrm{M_{\odot}}$ \\
        \hline
    \end{tabular}
\end{table}
}

KIC~9651065 is a binary system that has been extensively studied with phase modulation \citep{Murphy2015Deriving}. We used the 5 highest-amplitude frequencies in the model, while capturing the low-frequency variation with a GP as discussed in Sec.~\ref{sec:choice}. We show the derived time-delays of the system in Fig.~\ref{fig:9651065_td}, and the corresponding orbital fit in Table~\ref{tab:9651065}.

While forward-modeling of this system only offers a small improvement in the orbital parameters in this example, it provides a useful test-case to ensure the model functions correctly. We find that all parameters agree within the reported uncertainties, with the forward-model yielding slightly lower uncertainties. Since KIC~9651065 is a relatively long-period binary system, its eccentricity can be well described even with the subdividing method, since the chosen segment size of 10~d does not significantly reduce the orbital signal near periapsis.

\subsection{KIC 6780873: PB1, PM + RV}
\label{sec:6780873}

KIC\,6780873 is a short-period (9.15~d) binary with only one pulsating component (PB1). Previous modeling of this system using the subdivided approach determined a near-zero eccentricity \citep{Murphy2016Finding, Nemec2017Metalrich}. There is significant low-frequency variation in the amplitude spectrum of the star, which we modeled out using a high-pass filter for the subdividing method, and the GP for the forward-modeling method. We were able to further constrain the orbit with radial velocity measurements obtained from \citet{Nemec2017Metalrich}, following Eq.~\ref{eqn:radvel}. However, the radial velocity data are highly multi-modal and can be satisfied by a range of orbital parameters. Separate modeling revealed that the RVs alone are insufficient to fully constrain the orbit.

\begin{figure}
    \centering
    \includegraphics[width=\linewidth]{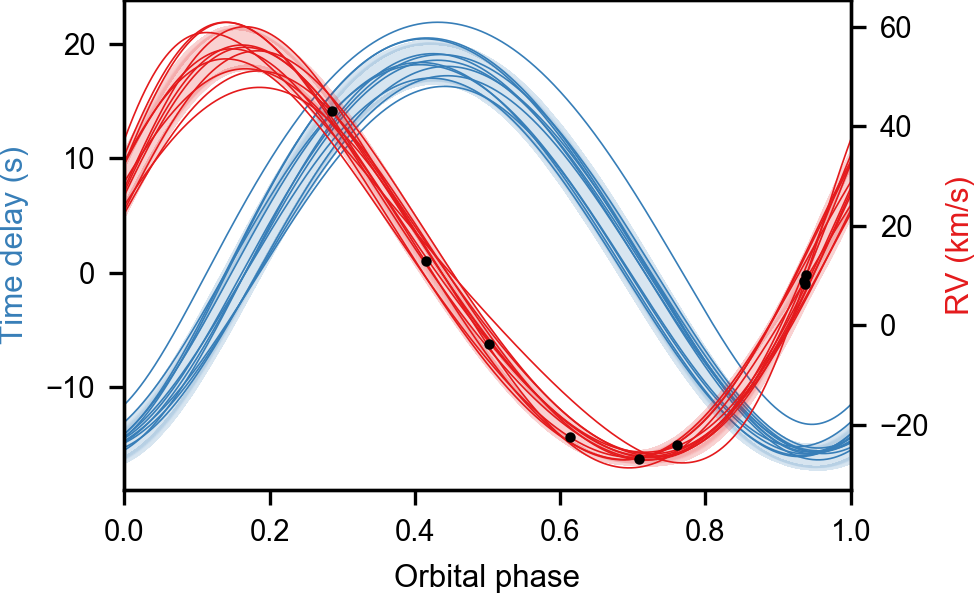}
    \caption{Radial velocity (red) and derived time-delay (blue) model fit from the forward-model for KIC~6780873. Shown are 10 randomly selected models from the trace after burn-in, along with the RV data (black points). The filled region shows the standard deviation of the posterior distribution. Reported uncertainties on the RVs are smaller than the data points.}
    \label{fig:6780873}
\end{figure}

We find that the forward-model both with and without RV data indicated a non-zero eccentricity (Table~\ref{tab:6780873}), in contrast to that found in previous papers. However, the subdividing method sampled with NUTS also determined a non-zero eccentricity. The discrepancy is potentially due to the Metropolis-Hastings sampler \citep{Metropolis1953Equation,Hastings1970Monte} used in previous papers. If the orbit is initially modeled as circular, then the phase of periapsis is undefined. Thus, every proposed step of the periapsis and eccentricity will be accepted under the Metropolis-Hastings algorithm and the sampler is more likely get caught in a local minimum. We show the derived time-delays and RV fit in Fig.~\ref{fig:6780873}.

The subdivided method yields a lower value for \asini\ than the forward-model. This is likely due to the non-zero eccentricity, which causes the orbital signal to be smeared and reduces the maximum observed time-delay. The forward-model alone also over-estimates the eccentricity when comparing to the combined RV data model, but correctly determines the remaining parameters within the given uncertainties.

{\renewcommand{\arraystretch}{1.4}%
\begin{table}
    \centering
    \caption{Fitted orbital parameters for KIC~6780873 through both the old and new methods.}
    \label{tab:6780873}
    \begin{tabular}{ccccc}
        \hline
           & Subdividing & {Forward} & Forward + RV & {Unit}  \\
        \hline
        $P_{\rm orb}$       & $9.159^{+0.003}_{-0.004}$     & $9.160^{+0.002}_{-0.003}$  &  $9.159^{+0.001}_{-0.001}$   &   d\\
        $\rm{a}_1\sin{i}/c$   & $14.4^{+1.1}_{-1.4}$          & $19.910^{+2.1}_{-1.7}$ &  $17.392^{+0.71}_{-0.62}$    &   s\\
        $e$                 & $0.15^{+0.15}_{-0.10}$        & $0.58^{+0.14}_{-0.12}$ &  $0.10^{+0.04}_{-0.03}$       &   \\
        $\varpi$            & $4.4^{+1.1}_{-1.9}$          & $1.2^{+0.2}_{-0.3}$ & $2.4^{+0.7}_{-0.6}$       &   rad    \\
        $\phi_p$            & $0.2^{+0.7}_{-0.7}$      & $0.26^{+0.2}_{-0.2}$ &  $0.38^{+0.19}_{-0.19}$      &     \\
%        $f(M)$   & $0.039^{+0.01}_{-0.01}$                & $0.101^{+0.03}_{-0.02}$  &  $0.067^{+0.009}_{-0.007}$ &  $\mathrm{M_{\odot}}$ \\
        \hline
    \end{tabular}
\end{table}
}

\subsection{KIC 10080943: PB2, PM+RV}
\label{sec:10080943}

\begin{figure}
    \centering
    \includegraphics[width=\linewidth]{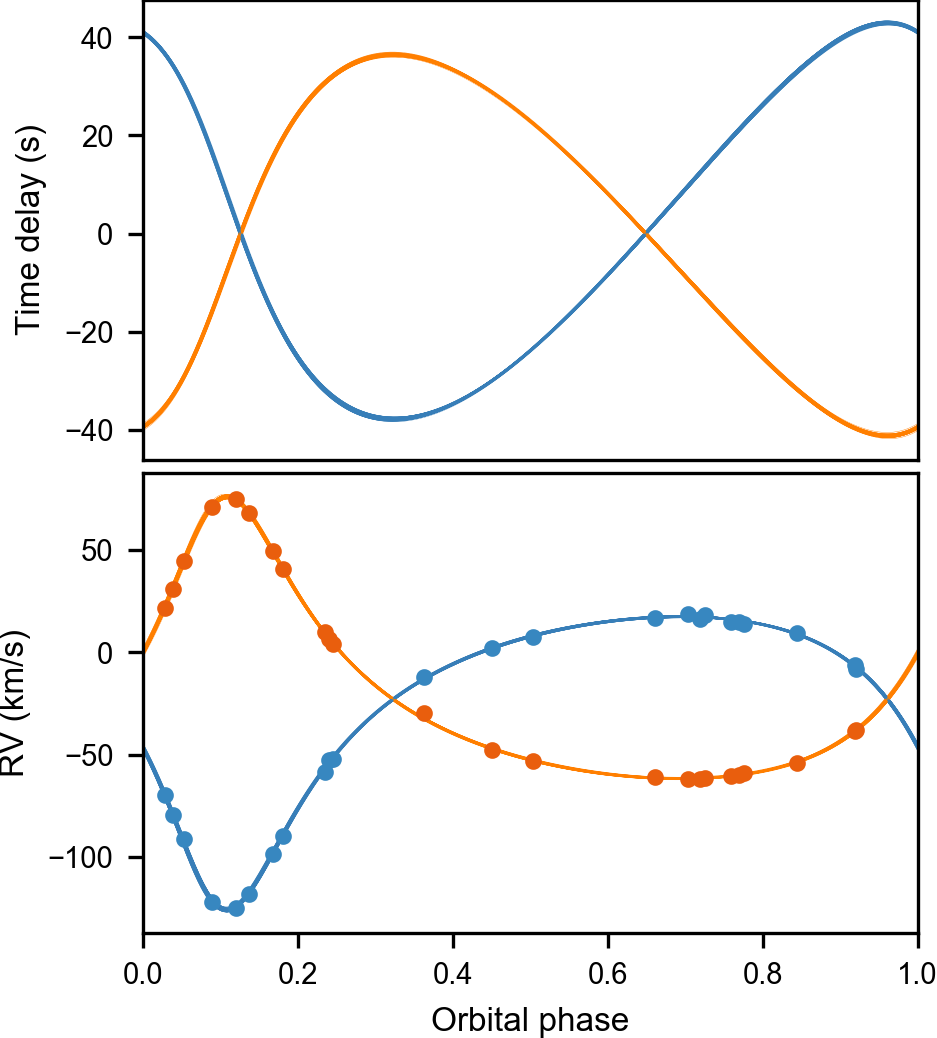}
    \caption{Top panel: Derived time delay model fit for both stars in the double pulsating (PB2) system, KIC~10080943, where the RV data has been used to constrain the solution. Shown are 10 randomly selected models from the posterior. The orange and blue lines correspond to the first and second stars respectively. Bottom panel: Simultaneous fit to the radial velocity data (blue and orange points) for the same stars.}
    \label{fig:10080943_td}
\end{figure}

{\renewcommand{\arraystretch}{1.4}%
\begin{table}
    \centering
    \caption{Fitted orbital parameters for KIC~10080943 for the forward-model and RV solutions. Note that a fit for the subdivided method has not been performed here as the model is not sensitive enough to the extracted time-delays alone. Here, $\gamma_V$ refers to the systemic velocity.}
    \label{tab:10080943}
    \begin{tabular}{cccc}
        \hline
           & {Forward} & {Forward + RV} & {Unit}  \\
        \hline
        $P_{\rm orb}$        &   $15.34^{+0.01}_{-0.01}$      &  $15.3363^{+0.0002}_{-0.0002}$    &d\\
        $\rm{a_1}\sin{i}/c$     &$49.0^{+7.3}_{-6.8}$ &  $44.94^{+0.18}_{-0.18}$     &s\\
        $\rm{a_2}\sin{i}/c$     & $41.9^{+6.1}_{-6.7}$  &  $43.16^{+0.15}_{-0.14}$     &s\\
        $e$                & $0.40^{+0.20}_{-0.23}$   &  $0.453^{+0.002}_{-0.002}$        &\\
        $\varpi$           &  $-0.11^{+0.64}_{-0.51}$   & $-0.270^{+0.008}_{-0.008}$        &rad    \\
        $\phi_p$             &  $0.51^{+0.61}_{-0.53}$    &  $0.593^{+0.008}_{-0.009}$     &  \\
        $\gamma_V$             &  --   &  $-22.86^{+0.13}_{-0.13}$     &km/s  \\
        % $f(M)$         &$0.74^{+0.46}_{-0.33}$ &  $0.414^{+0.005}_{-0.005}$ &  $\mathrm{M_{\odot}}$ \\
        \hline
    \end{tabular}
\end{table}
}
KIC\,10080943 is a 15-d binary system composed of two $\gamma$ Doradus and $\delta$~Sct hybrid stars \citep{Schmid2015KIC, Keen2015KIC} which lies at the edge of the detectability limit for the subdividing method \citep{Compton2016Binary}. In this system, both stars show time-delays appearing in anti-phase (PB2), analogous to the double-lined spectroscopic binaries (SB2). KIC~10080943 previously only had an orbital solution derived when fitting either the RVs individually or with a combination of RVs and PM. We found that the forward-model successfully captures the orbital properties of the system without any RV data, albeit with much greater uncertainty. 

We show our derived orbital elements in Table~\ref{tab:10080943}. While the projected semi-major axis of star 1 is overestimated without RV data, all of the remaining parameters agree within the relatively large uncertainty. The system is moderately eccentric with a short orbital period, a solution that the forward-model can correctly extrapolate.

\subsection{KIC 7668647: a sdB+WD binary}
\label{sec:sdb_wd}

Phase modulation is not limited to $\delta$~Sct stars. Indeed, other variable star pulsations have been used as the `clock' for identifying binarity to varying degrees of success (see \citealt{Hermes2018Timing} for a full discussion). One of these systems, KIC~7668647, is one of the first stars for which the stellar pulsations were used to determine orbital properties \citep{Telting2012Three, Telting2014KIC}.

KIC~7668647 is a binary composed of a subdwarf B star (sdB) and a white dwarf (WD). \citet{Telting2014KIC} demonstrated that simultaneously fitting all pulsation frequencies in the power spectrum yielded an \asini\ in good agreement with the value derived from radial velocities. Their approach only allowed the time-delay -- the maximum projected light travel time across the orbit -- and time zero-point to be free parameters, with the other orbital elements fixed to their known values obtained from spectroscopy.

There are over 10 quarters of \textit{Kepler} short cadence data available for KIC~7668647. Given the extremely short orbital period, we used only quarters 14 through 17 of the data to reduce computation time. The sdB star in this system oscillates in high-frequency g-modes, in a similar frequency range of the p-mode pulsations found in $\delta$~Sct stars \citep{Charpinet2013GMode}. We used these pulsations to obtain the full orbital parameters. Since the GP described in Sec.~\ref{sec:choice} was designed to remove such variation, it was not applied to this model.

We found that several of the strongest pulsation modes in KIC~7668647 undergo significant intrinsic frequency modulation not caused by binarity. As a result, if these pulsation modes are included then the resulting fit is skewed since the model is weighted by the amplitude of the modes. We identify such modes following Sec.~\ref{sec:choice}.

Using the modes that do not undergo strong intrinsic modulation, we provide the fitted parameters in Table~\ref{tab:7668647} noting that we could not determine a solution with the subdividing method even with a fixed orbital period and zero eccentricity. The orbital period and projected semi-major axis are comparable to those obtained by \cite{Telting2014KIC} using radial velocities, within the given uncertainties. However, the eccentricity is poorly constrained, which we attribute to the incoherency of the g-mode pulsations in subdwarf B stars.

{\renewcommand{\arraystretch}{1.5}%
\begin{table}
    \centering
    \caption{Fitted orbital parameters for KIC~7668647 for the forward-model compared against spectroscopic values \citep{Telting2014KIC}. Note that a fit for the subdivided method has not been performed here as the model is not sensitive enough. $\dagger$Although a non-zero eccentricity was obtained, the model was fixed to a circular orbit and thus the periapsis was undefined.}
    \label{tab:7668647}
    \begin{tabular}{cccc}
        \hline
           & {Forward} & {\citealt{Telting2014KIC} (RV)} & {Unit}  \\
        \hline
        $P_{\rm orb}$        &   $14.10^{+0.1}_{-0.1}$      &  $14.1742\pm0.0042$    &d\\
        $\rm{a}_1\sin{i}/c$     &$53.3^{+7.9}_{-8.1}$ &  $50.8\pm2.5$     &s\\
        $e$                & $0.4^{+0.3}_{-0.3}$   &  $0.056\pm0.006$        &\\
        $\varpi$           &  $-1.1^{+1.2}_{-2.9}$   & $[0]^\dagger$        &rad    \\
        $\phi_p$             &  $0.9^{+2.0}_{-1.2}$    &  $0.593^{+0.008}_{-0.009}$      \\
        % $f(M)$         &$0.74^{+0.46}_{-0.33}$ &  $0.414^{+0.005}_{-0.005}$ &  $\mathrm{M_{\odot}}$ \\
        \hline
    \end{tabular}
\end{table}
}

\section{Detection limits}
\label{sec:detection_limits}

\begin{figure*}
    \centering
    \includegraphics[width=\linewidth]{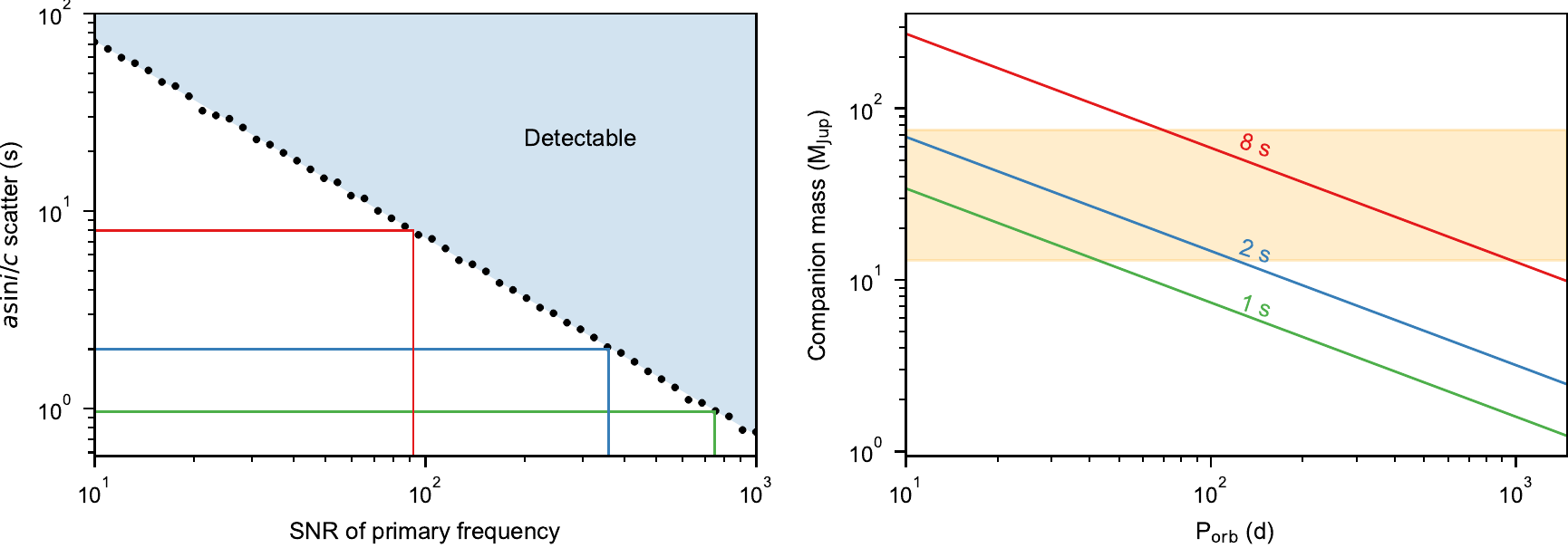}
    \caption{\textit{Left}: Results of Monte Carlo simulations for the scatter in \asini\ as a function of the SNR of the primary frequency (black points). The shaded blue region indicates areas where the forward-model should be capable of identifying binarity. The red, blue, and green lines indicate the 16th, 50th, and 84th percentile values of $\delta$~Sct star pulsation SNR (in \textit{Kepler}) respectively for both plots. \textit{Right}: Companion Jovian masses corresponding to the expected detection limits in \asini\ for objects orbiting an ideal pulsating $\delta$~Sct star, updated from Figure~9 of \citep{Murphy2016Finding}. We assume the primary to be 1.8~$\mathrm{M_{\odot}}$ and $\sin{i}=1$. The yellow shaded area represents the canonical mass range of brown dwarfs.}
    \label{fig:planet_mass_compare}
\end{figure*}

\begin{figure}
    \centering
    \includegraphics[width=\linewidth]{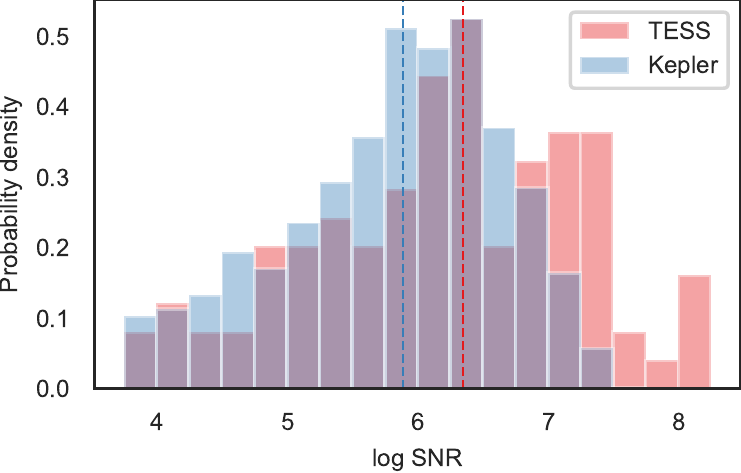}
    \caption{Distribution of the SNR of pulsations for the known $\delta$~Sct stars in the \textit{Kepler} \citep{Murphy2019Gaiaderived} and \textit{TESS} (this work) fields. The red and blue lines correspond to the median value of the distribution.
    }
    \label{fig:dsct_snr}
\end{figure}

We now seek to characterize the detection limits of the new method and compare with the previous approach. To achieve this, we generated \textit{Kepler} time-series which simulate binary motion. The simulated light curves were a linear combination of two components: a set of pulsation modes and a white noise term (Eq.~\ref{eq:luminosity}). Following \cite{Compton2016Binary}, the SNR of the pulsations was adjusted by injecting white noise $\epsilon$ at each time $t_n$
\begin{eqnarray}
\epsilon(t_n) = \mathcal{N}(0,\sigma_{rms}),
\end{eqnarray}
where
\begin{eqnarray}
\sigma_{rms} = \dfrac{A}{\sqrt{\frac{\pi}{N}} \rm (SNR)},
\label{eq:snr}
\end{eqnarray}
where $A$ is the oscillation amplitude of the strongest mode and $N$ is the number of data points. Binary motion was simulated by adding a time-dependent phase shift to each pulsation mode, following Eq.~\ref{eq:delay}.

\subsection{Simulations without binarity}

\begin{figure*}
    \centering
    \includegraphics{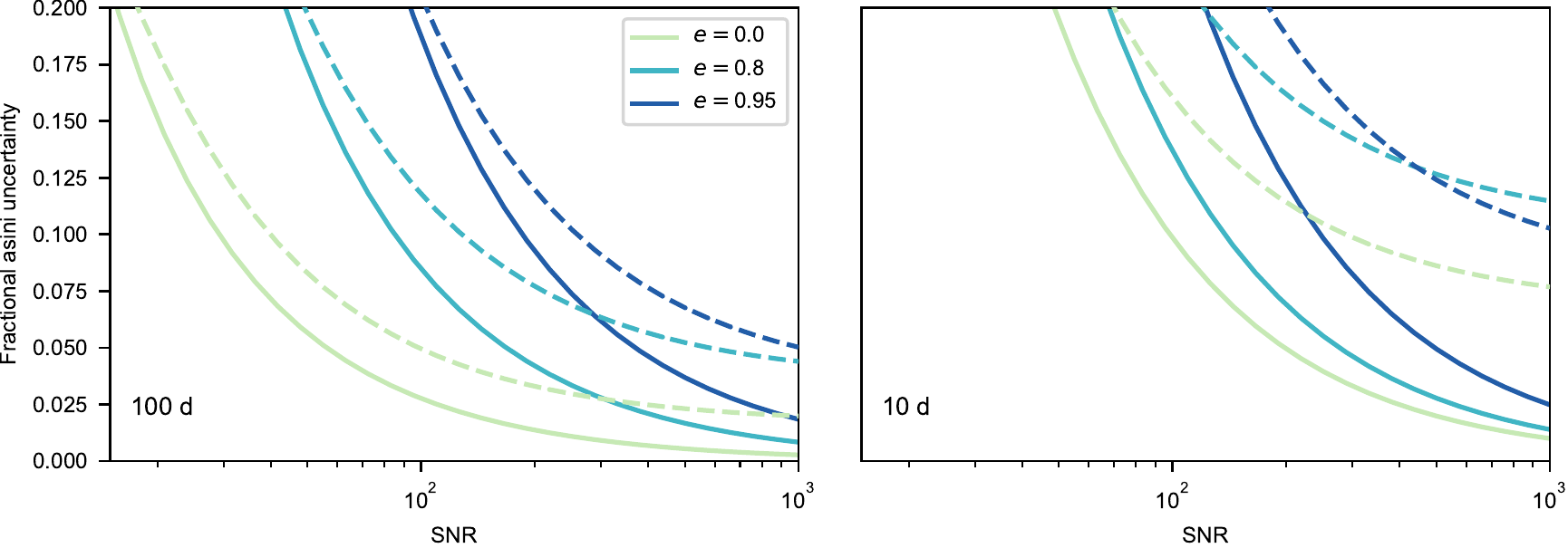}
    \caption{Comparison of fractional \asini\ uncertainties for a synthetic binary system at different SNR and eccentricities. The solid and dashed lines represent the forward and subdivided models respectively.}
    \label{fig:snr_comparison}
\end{figure*}

We first considered the absence of binarity in the light curves. That is, what is the inferred \asini\ when no binary motion is included? We used a single-mode $\delta$~Sct star pulsating at 30~d$^{-1}$, and generated a set of light curves of SNR between 10 and 1000, following Eq.~\ref{eq:snr}. For each SNR, we created 5000 randomly generated light curves with no binary signal which were fitted to Eq.~\ref{eqn:lightcurve}, letting the maximum time-delay be the only free parameter.

We show the standard deviation of the fitted maximum time-delay for each SNR in Fig.~\ref{fig:planet_mass_compare}. The standard deviation is a useful indicator of the minimum detectability limit of the forward-model as a function of the SNR of the primary frequency and is similar to the scatter used by \cite{Compton2016Binary} for probing detectability of the subdividing method. A linear fit to the SNR vs. \asini\ scatter yields the relation,

\begin{eqnarray}
    \ln{(\sigma_{a\sin{i}})} = -1.0048 \ln{(\rm{SNR})} + 6.6173.
\end{eqnarray}

To place these values in the context of known $\delta$~Sct stars, we calculated the SNR of the strongest pulsation peak within the \textit{Kepler} $\delta$~Sct catalogue of \cite{Murphy2019Gaiaderived} (Fig.~\ref{fig:dsct_snr}). The 16th, 50th, and 84th percentiles in the distribution correspond to SNRs of 92, 358, and 750 respectively. We used these values as measures of the typical pulsation SNR for $\delta$~Sct stars and found that a pulsation SNR of 358 for the median $\delta$~Sct leads to a detectability limit of 2~s, which is sufficient to identify a wide range of brown dwarf companion masses provided the orbital period is greater than 100~d. In the best case (SNR > 1000), 1~$M_{\rm Jup}$ planetary companions are detectable with periods exceeding 1000~d. Around $2000$ $\delta$~Sct stars have been observed in long-cadence in the \textit{Kepler} nominal mission \citep{Murphy2019Gaiaderived}. Given that planetary signals are difficult to detect using RV around A-type stars due to Doppler broadening, our approach provides a novel method of searching for Jupiter mass exoplanets.

For \textit{TESS}, stars that lie within the continuous viewing zone (CVZ) in the Southern Hemisphere have one year of continuous photometry. These observations are currently being repeated for stars in the Northern Hemisphere, providing a long baseline of observational data from which time-delays can be extracted. An extensive search for orbital companions with pulsation timing has not yet been performed on these stars in the CVZ. However, we instead provide a rough estimate of the yield of $\delta$~Sct stars. We performed a query of all stars that have at least 10 sectors of data in \textit{TESS}, and lie within a temperature range of 6500 to 10 000~K according to the Tess Input Catalogue (TIC; \citealt{Stassun2018TESS}). Following \cite{Murphy2019Gaiaderived}, we searched for $\delta$~Sct stars using the skewness of the amplitude spectrum as an indicator of variability, confirming the pulsations manually. Of the 794 \textit{TESS} stars, we identified 95 $\delta$~Sct stars, leading to an occurrence rate of 12\% within the CVZ. We calculated the median value of the SNRs of these stars to be 504 (Fig.~\ref{fig:snr_comparison}), significantly higher than the median value of 358 in \textit{Kepler}. This points to a median detectability limit of around 1.5~s. If \textit{TESS} proceeds to return to observations in the Southern Hemisphere after the nominal mission, the base-line of the current CVZ photometry will be extended by another year, allowing for systems with orbital periods greater than 2 years to be resolved.

Given that significantly more stars in the CVZ have been observed in the long-cadence (30 min sampling) over short-cadence (2-min) mode with \textit{TESS}, the yield (and subsequently, SNR) should change significantly. We leave these results to a future paper once the extracted light curves are made available.

\subsection{Comparison with the subdividing method}

We next considered the effects of eccentricity on both models by generating a grid of artificial $\delta$~Sct light curves with pulsation SNR logarithmically spaced between 10 and 1000. For each SNR, we injected a binary signal with \asini\ of 100~s and eccentricity of 0, 0.5, and 0.99 respectively. We considered both a short period binary (10~d) and a longer period binary (100~d). We optimized for both the subdividing and forward-models and recorded the returned \asini\ at each grid point. For the subdividing method, we chose the segment size as $P_{\rm orb} / 3$ to fully capture the orbital signal. We compare the fractional uncertainty for both methods in Fig.~\ref{fig:snr_comparison}. 

We found that the forward-model performs better in both the long and short orbital period systems. For longer periods binaries (>100~d), the fractional uncertainty for both methods is quite close, with the subdividing method yielding less accurate values due to smearing of the orbital signal with increasing eccentricity. For the shorter period binary, forward-modeling performs significantly better in obtaining the true value of \asini. The downside of the new approach is that it is significantly slower than the subdividing method as a result of individually fitting each observation in the light curve. Thus, for longer-period binaries with low eccentricity, it is more efficient to use the subdividing method to obtain orbital parameters. For more sensitive cases, such as short-period binaries or exoplanets, the forward-model approach should be used instead.

As mentioned in Sec.~\ref{sec:choice}, the pulsation frequencies themselves determine whether an orbital signal can be observed. When the ratio of \asini\ to the pulsation frequency is large, the pulsations are more sensitive to the orbital variations. This is illustrated in Figure~2 of \cite{Compton2016Binary}, alongside the effect of the signal-to-noise of the pulsations mode. Subdividing the light curve is problematic when there are closely spaced peaks in the amplitude spectrum, since such peaks are unresolved in the short subdivisions, leading to a signal at the beat frequency of the modes \citep{Murphy2016Finding}. Since our forward-model approach involves no segmentation of the light curve it is instead limited by the cadence of the observations and the SNR of the pulsations \citep{Shahram2005Resolvability}.

\subsection{A simulated planet in the Kepler field}

As a final test of detection limits, we performed a hare-and-hounds exercise with a range of sub-stellar mass companions in the \textit{Kepler} field. We considered four separate systems with a planet of 1, 2, 5, and 10 Jupiter masses in a circular orbit around a $\delta$~Sct star of 1.8~$M_\odot$. The orbital period was fixed to 1000~d so that the orbits only differed by their respective \asini. The SNR of the highest amplitude pulsation frequency was set to the median value for a $\delta$~Sct star in the \textit{Kepler} field: 358, corresponding to the green line in Fig.~\ref{fig:planet_mass_compare}.

{\renewcommand{\arraystretch}{1.4}%
\begin{table}
    \centering
    \caption{Fitted masses for a simulation of a range of Jupiter mass planets orbiting a 1.5~$M_\odot$ $\delta$~Sct star at a SNR of 358. A solid line indicates models which did not converge to a solution.}
    \label{tab:jmass_fit}
    \begin{tabular}{ccccc}
        \hline
       {$P_{\rm orb}$}  &    {1 $M_{\rm Jup}$}   &  {2 $M_{\rm Jup}$}   &  {5 $M_{\rm Jup}$} &   {10 $M_{\rm Jup}$} \\
        \hline
        $1000$      &   --- &  ---   &   $4.76^{+0.80}_{-0.80}$   &   $9.87^{+0.77}_{-0.73}$  \\
        \hline
    \end{tabular}
\end{table}
}
For simplicity, we chose to have three modes with typical frequencies from the range of $\delta$~Sct stars, 30, 40, and 50 d$^{-1}$, and generated a light curve with binary orbital motion injected. We attempted to fit these systems using the forward-model, and show our results in Table~\ref{tab:jmass_fit}. As expected, the forward-model can not identify a 1 or 2~$M_{\rm Jup}$ planet at an orbital period of 1000~d, since the corresponding \asini\ is 0.63~s and 1.3~s respectively, which is below the expected minimum detectable signal determined in Sec.~\ref{sec:detection_limits}. On the other hand, we recovered the remaining planets within uncertainties.

%\subsection{Low frequency noise}

%\textbf{To ensure robustness against low frequency noise, we re-analyzed the SAP flux data of KIC~9651065. We used the same model and priors as in Sec.~\ref{sec:9651065} without a GP model, and found that the obtained values are almost identical to those with the PDCSAP data. As another test, we generated a synthetic system with included white (uniformly random) and significant red (low frequency) noise. We attempted to recover the known orbital parameters using the forward-model with and without a GP model, and found that both systems converged to the correct solution within the uncertainties. We summarise these comparisons in Table.~\ref{tab:low_freq_comparison}. As expected, the model with the GP is slightly better in obtaining the correct values, due to the fact that the GP was designed to account for such low frequency variations, whether they are pulsations or otherwise.}

\section{Discussion}

\begin{figure}
    \centering
    \includegraphics{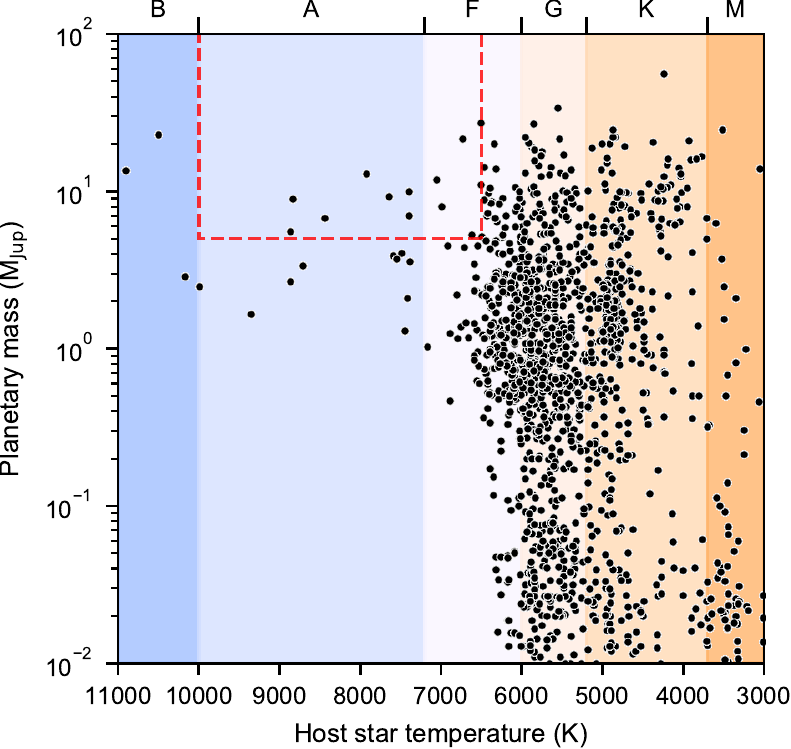}
        \caption{All confirmed planetary systems from the \textit{NASA} exoplanet archive (black points). The red dashed box indicates the region in which the forward-modeling approach is typically sensitive within the \textit{Kepler} data.}
    \label{fig:all_koi}
\end{figure}

In this paper, we made several key assumptions about the data. The first was that we interpreted orbital signals as originating from a binary system, although each system may be a triple or higher-order multiple. If there are more than two components in the system, then the detection of orbital motion depends upon the configuration of the system. In a triple system, it is common for two of the stars form a close binary and the tertiary orbits the pair at a larger distance: a hierarchical triple system. If the pulsating component lies in the close binary, the observed time delays will be a superposition of the short- and long-period orbits. If the pulsating component is the tertiary, the observed time delays will show a single orbit, where the phases of the oscillation modes are instead modulated by the sum of the masses of the inner pair.

Another assumption made was that the oscillation modes are nearly or completely coherent throughout the \textit{Kepler} mission lifetime. Departures from coherence will influence whether an orbital signal can be observed. Although intrinsic and extrinsic (i.e., due to orbital motion) frequency or phase modulation are indistinguishable for an individual pulsation mode, orbital modulation influences all oscillation modes equally. As a result, intrinsic variations caused by incoherence can be disentangled from orbital modulation as long as the orbital signal is sufficiently strong and there are enough excited oscillation modes whose modulations are in mutual agreement. Another case where incoherent oscillation modes can be used for PM is when the orbital period is short enough that the oscillation remains stable for several orbital phases, as was the case for KIC~7668647 (Sec.~\ref{sec:sdb_wd}).

For all examples discussed in Sec.~\ref{sec:examples}, we made use of the PDCSAP photometry provided by the \textit{Kepler} mission. One of the features of this pipeline is a high-pass filter, which removes long period variations such as instrumental artefacts. As a result, such low frequency noise does not impact the final values obtained by the forward-model. If low frequency signals were present, in the form of g-mode oscillations or simply instrumental artefacts, then the GP model is sufficient to capture the slow variations.

In general, the forward-modeling approach seems most useful for verifying planetary transits by ruling out brown dwarf or stellar companions to early-type stars where RVs are not available. A- and F-type stars on the main sequence are generally omitted from RV surveys because of their rapid rotation and broad spectral lines, resulting in few confirmed exoplanets around hotter stars. Thus, there is a significant gap in population statistics for exoplanet host stars above 6500 K. Phase modulation is uniquely situated to complete these statistics for higher mass companions (Fig.~\ref{fig:all_koi}). For stellar companions, this technique is useful for highly eccentric and short-period binary systems, where subdividing the light curve introduces smearing of the orbital signal near periapsis.

\section{Conclusion}
We have formulated a new approach to modeling light travel time variations in a binary orbit by simultaneously fitting all points in the light curve, avoiding the need to subdivide the light curve into segments. Our approach covers a wider range of systems than the previous method, which underestimates the projected semi-major axis for short-period and eccentric binaries due to smearing of the orbital signal. We have shown that this approach significantly decreases the required SNR of the oscillation modes.

Under ideal conditions, we found that the detection limit for a typical \textit{Kepler} $\delta$~Sct star is a projected semi-major axis of around 2~s. For stars pulsating with higher SNR, 1~$M_{\rm Jup}$ exoplanets should be detectable as long as their orbital period is greater than 1000~d, and the $\delta$~Sct oscillation is coherent over the orbital phase.

For space-based missions with long-baseline photometry, PM represents a unique method for confirming transiting exoplanets around the A/F type stars. In particular, the \textit{PLATO} mission will perform high precision uninterrupted photometric observations of bright stars \citep{Rauer2014PLATO}. While the primary science goal is the detection of planetary transits, measurements of pulsations in $\delta$~Sct stars will enable the use of PM.

\section*{Acknowledgements}
We are thankful to both the \textit{Kepler} and \textit{TESS} teams for such fantastic data. We thank the anonymous referee, whose comments greatly improved the quality of this manuscript. DRH gratefully acknowledges the support of the Australian Government Research Training Program (AGRTP) and University of Sydney Merit Award scholarships. This research has been supported by the Australian Government through the Australian Research Council DECRA grant number DE180101104, and by the Simons Foundation. This work was performed in part under contract with the Jet Propulsion Laboratory (JPL) funded by NASA through the Sagan Fellowship Program executed by the NASA Exoplanet Science Institute. This research has made use of the NASA Exoplanet Archive, which is operated by the California Institute of Technology, under contract with the National Aeronautics and Space Administration under the Exoplanet Exploration Program.

\vspace{5mm}

\software{astropy \citep{astropy:2013, astropy:2018},  
          Lightkurve \citep{GeertBarentsen2019KeplerGO},
          Matplotlib \citep{ThomasACaswell2019Matplotlib},
          Scipy \citep{Jones2001SciPy},
          PyMC3 \citep{Salvatier2016Probabilistic},
         Numpy \citep{Oliphant2015Guide}
          }

\bibliography{library}{}
\bibliographystyle{aasjournal}

%% This command is needed to show the entire author+affiliation list when
%% the collaboration and author truncation commands are used.  It has to
%% go at the end of the manuscript.
%\allauthors

%% Include this line if you are using the \added, \replaced, \deleted
%% commands to see a summary list of all changes at the end of the article.
%\listofchanges

\end{document}